# Quantitative Analysis of Composition and Contamination of Atomically Thin Materials by Recoil-Projectile Coincidence in Ion Transmission

Carolin Frank[1,2*], Kevin Vomschee[1], Tuan Thien Tran[1], Barbara Maria Mayer[1], Radek Holeňák[3], Yossarian Liebsch[2], E. Harriet Åhlgren[1,4], Marika Schleberger[2], and Daniel Primetzhofer[1,3]

[1]*Materials Physics, Department of Physics and Astronomy, Uppsala University, Uppsala, Sweden*
[2]*Faculty of Physics and CENIDE, University of Duisburg-Essen, Duisburg, Germany*
[3]*Tandem Laboratory, Uppsala University, Uppsala, Sweden*
[4]*Department of Physics, University of Helsinki, Helsinki, Finland*

*Corresponding author: carolin.frank@physics.uu.se

## Abstract

Surface contamination strongly affects the intrinsic properties of nanoscale materials, making its reliable identification and quantification crucial for both accurate experimental interpretation and nanofabrication. Although scanning transmission electron microscopy can resolve contaminants at atomic resolution within nanometer-scale regions, it cannot easily provide a quantitative, large-area contamination measure. Here, we introduce a minimally destructive recoil-projectile coincidence method for ion transmission experiments that enables element-specific identification and quantification of surface contaminants with isotopic resolution. We demonstrate this approach by comparing self-supporting graphene samples prepared using either a polymethylmethacrylate (PMMA)-based or a PMMA-free transfer process. Carbon and hydrogen are identified as the dominant surface contaminants. PMMA-free transferred graphene exhibits the lowest native contamination levels. Following in-situ thermal annealing at 400 °C for 1 h, the measured carbon areal density approaches the value expected for atomically clean single-layer graphene within the experimental uncertainty, while hydrogen coverage is strongly reduced. Unlike PMMA-transferred graphene, which rapidly recontaminates after annealing, PMMA-free transferred graphene remains nearly contamination-free for at least 140 min under ultra-high vacuum conditions ($p_{base}$ = 2x10$^{-8}$ mbar). Beyond graphene, the presented method establishes a quantitative characterization platform for ultrathin materials, enabling studies of surface cleanliness, adsorption, implantation and surface interaction dynamics in such systems.

Keywords: graphene, surface contamination, ion beam analysis, time-of-flight medium energy ion scattering, recoil-projectile coincidence, ultrathin materials

## 1. Introduction

Surface contamination is a persistent challenge in both surface science and nanofabrication, as it can significantly alter the intrinsic properties of surfaces[1–3]. This issue is particularly critical for two-dimensional (2D) materials, which consist entirely of surface atoms. Consequently, their properties, including doping level, carrier mobility, strain, thermal conductivity and wettability, are highly sensitive to contamination[2,4–11]. Contaminants are introduced already during sample preparation and subsequent handling, particularly when samples are exposed to ambient conditions, such as after mechanical exfoliation or following chemical vapor deposition (CVD) growth[4,12,13]. Freshly prepared sample surfaces are therefore rapidly covered by airborne hydrocarbons, water and oxygen-containing adsorbates, while additional silicon-based impurities may be introduced depending on the synthesis route[1,3,14,15]. Furthermore, transfer procedures required for the fabrication of suspended or freestanding 2D materials introduce additional contamination[16–19]. For graphene, transfer processes commonly employ polymethylmethacrylate (PMMA) for mechanical stabilization[12,20]; however, PMMA residues are difficult to remove completely, even by solvent treatment with acetone[11,21]. In contrast, PMMA-free transfer processes have been shown to significantly reduce surface contamination levels of graphene[22,23].

Since the discovery of graphene in 2004[24], numerous ex-situ and in-situ cleaning methods have been developed, including laser- and plasma-based treatments as well as thermal annealing[18,21,25–27]. An ideal cleaning process would produce large atomically clean surface areas with minimal effort while preserving the intrinsic properties of the sample and avoiding introduction of strain or doping. Achieving these requirements, however, remains challenging. While laser-based cleaning methods are often limited to the accessible laser spot size and plasma-based cleaning methods are strongly dependent on process-specific parameters such as plasma energy and exposure time, thermal annealing under ultra-high vacuum (UHV) conditions has been shown to be an effective route toward obtaining large atomically clean graphene surfaces[25,27,28].

Using atomically resolved scanning transmission electron microscopy (STEM) images, Irschik et al. recently reported that thermal annealing above 400°C in UHV efficiently removes nearly all airborne hydrocarbon and polymer contamination from both graphene and hexagonal boron nitride (h-BN)[27]. While the development of effective cleaning procedures is essential, reliable quantification of the remaining surface contamination is crucial for the accurate characterization of nanoscale materials and the correct interpretation of experimental results. However, despite the capability of microscopy techniques such as STEM to resolve contaminants at the atomic scale within nanometer-scale sample regions, establishing a reliable quantitative measure of the overall surface contamination remains a significant challenge [4,29]. Furthermore, microscopy techniques generally provide only limited elemental sensitivity for physisorbed light species such as hydrogen and may influence the surface state during imaging[29].

In contrast to microscopy techniques that provide highly localized information, non-destructive ion beam analysis techniques such as Rutherford Backscattering Spectrometry (RBS), Elastic Recoil Detection (ERDA) or Nuclear Reaction Analysis (NRA) enable the quantitative determination of element concentrations, depth profiles and thicknesses over comparably large sample areas. However, for ultrathin films, transmission recoil analysis with projectile energies in the keV range provides enhanced depth resolution and surface sensitivity[30]. While graphene purity has been investigated in backscattering geometry using low energy ion scattering (LEIS)[31], the application of transmission recoil analysis of 2D materials remains largely unexplored.

Building on the transmission recoil analysis approach for determining the average thickness of a contaminated graphene sample using medium energy ion scattering (MEIS) reported in reference [32], we developed a minimally destructive method for the quantitative assessment of surface contamination. By detecting recoils in coincidence with the transmitted projectile, individual contamination species can be unambiguously identified and quantified with a precision primarily limited by counting statistics, while the absolute accuracy depends on the effective scattering cross sections and detector acceptance. In particular, our recoil-projectile coincidence approach enables separation of light contamination species such as carbon and oxygen and provides isotopic resolution. The capabilities of our method are demonstrated using 40 keV Xe ions transmitted through self-supporting single-layer graphene (SLG) on TEM grids at the Time-of-Flight Medium Energy Ion Scattering (ToF-MEIS) setup at Uppsala University.

In this article, we introduce this method to quantify surface contamination and apply it to self-supporting SLG to investigate

- the influence of both annealing time and temperature on contamination removal,
- the impact of PMMA-based and PMMA-free transfer processes on graphene cleanliness, and
- the recontamination behaviour of differently prepared samples subsequent to thermal annealing in UHV.

Our results show that surface contamination of graphene consists predominantly of carbon and hydrogen. The measurements reveal pronounced differences in both contamination removal and post-annealing recontamination depending on the transfer route: Among the investigated samples, PMMA-free transferred graphene, hereafter referred to as Flattened graphene, exhibits the lowest native levels of both carbon and hydrogen contamination. Furthermore, we demonstrate that thermal annealing at 400 °C for 1 h reduces the measured C and H coverages to values consistent with nearly atomically clean graphene within the sensitivity of the method, independent of the sample transfer process. However, while carbon and hydrogen contamination rapidly readsorb on PMMA-transferred samples, PMMA-free transferred Flattened graphene remains nearly contamination-free for at least 140 minutes after annealing under UHV conditions at a base pressure of $2 \times 10^{-8}$ mbar.

The present approach provides a quantitative route to determine contamination and cleaning efficiency in atomically thin targets, enabling a more reliable correlation between surface cleanliness and measured material properties.

## 2. Results

First, we analyze how thermal annealing under UHV conditions affects the cleanliness of the three graphene samples described in Section 5.

### 2.1 Effect of annealing temperature and duration on PMMA-free transferred graphene

We first examine the effect of annealing temperature and duration on the reduction of surface contamination. Figure 1a) shows energy spectra of 40 keV Xe projectiles transmitted through self-supporting Flattened graphene measured by ToF-MEIS. In all measurements, the coincidence condition described in Section 5 was applied.

The corresponding hydrogen and carbon coverages, expressed in monolayers equivalents, are shown in Figure 1b) for each spectrum. The error bars represent the relative uncertainty arising from counting statistics. The ideal value of one carbon monolayer expected for clean single-layer graphene is indicated by a dashed line.

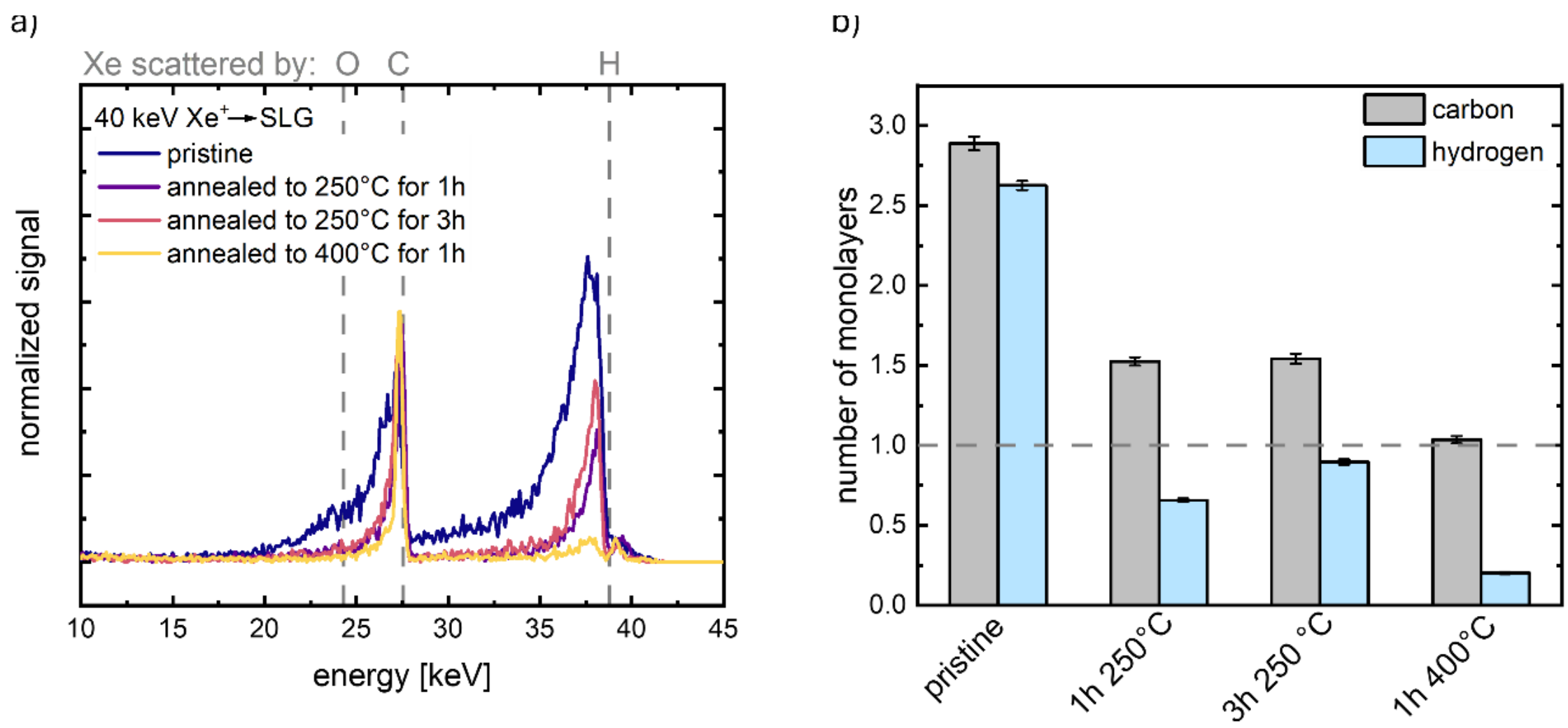

Figure 1: a) Measured signal of detected Xe projectiles as a function of recorded energy after transmission of 40 keV Xe ions through single-layer Flattened graphene. The spectra are normalized to the number of Xe projectiles passed through graphene, respectively. b) Corresponding number of C and H monolayers of the non-annealed sample, after annealing at 250 °C for 1 h, after annealing at 250 °C for 3 h and after annealing at 400 °C for 1 h. The error bars correspond to the relative uncertainty arising from counting statistics.

In Figure 1a), the recorded energy spectra are shown for the pristine sample (blue), after annealing at 250 °C for 1 h (purple), after annealing at 250 °C for 3 h (pink) and after annealing at 400 °C for 1 h (yellow). All spectra are normalized to the number of Xe projectiles transmitted through the Flattened graphene sample, respectively. Post-annealing measurements were performed after the sample temperature decreased to 40 °C.

In each spectrum, two well-separable features are present, corresponding to Xe ions scattered by hydrogen atoms and carbon atoms, respectively. The corresponding energies for scattering from hydrogen (38.78 keV), carbon (27.53 keV), and oxygen (24.29 keV), obtained by classical mechanics, are indicated by dashed line. The signals with very low intensity around 39 keV are attributed to “false coincidences”, i.e., events in which detected particles fall within the selected time-of-flight range and fulfill the coincidence condition but did not contribute to the scattering process.

The energy spectra shown in Figure 1a) strongly differ from each other in terms of peak intensity and width. For the pristine, non-annealed sample, the intensity of the hydrogen-related signal is significantly higher than that of the carbon-related peak. In addition, a shoulder at approximately 24 keV is indicated, corresponding to Xe projectiles scattered by oxygen atoms. The corresponding contaminations shown in Figure 1b) are 2.89 ± 0.04 monolayers of C and 2.63 ± 0.09 monolayers of hydrogen. After annealing at 250 °C for 1 h, the hydrogen-related signal is reduced below the carbon-related peak intensity. Also, the peak width of both peaks is strongly reduced compared to the spectrum of the pristine sample. As shown in Figure 1b), the corresponding amounts of contamination are 1.53 ± 0.03 monolayers of C and 0.66 ± 0.02 monolayers of hydrogen.

Following annealing at 250 °C for 3 h, the hydrogen signal increases again, whereas the carbon peak intensity and width remain comparable to the 1 h case, resulting in 1.54 ± 0.03 monolayers of carbon and 0.9 ± 0.03 monolayers of hydrogen.

After annealing at 400 °C for 1 h, the hydrogen signal is reduced nearly to zero, while the width of the carbon peak is significantly reduced compared to the lower-temperature annealing steps. This corresponds to 1.04 ± 0.02 monolayers of carbon and 0.2 ± 0.01 monolayers of hydrogen, i.e. 0.1 monolayers of hydrogen per graphene surface.

### 2.2 Influence of the transfer process on initial contamination and cleaning efficiency

To study the effect of the sample transfer process on graphene cleanliness, ToF-MEIS measurements were performed on Flattened graphene, Easy Transfer graphene and Graphenea graphene after transfer onto TEM grids, as described in Section 5. Each sample was measured before annealing and after annealing at 400 °C for 1 h. Measurements of the annealed samples were carried out after cooling to 40 °C. Measurements of the annealed samples were performed 2 h 50 min after annealing for Flattened graphene, 2 h 4 min for Graphenea graphene, and 2 h 16 min for Easy Transfer graphene.

In Figure 2a), we show the time-of-flight spectra recorded before and after annealing for Flattened graphene (blue), Easy Transfer graphene (purple) and Graphenea graphene (orange). In each panel, the measured signal is normalized to the detected number of transmitted Xe projectiles. The flight time of the direct 40 keV Xe ion beam (1185.5 ns) is indicated by a dashed line. All time-of-flight spectra are shown in separate figures in the Supplementary Information[33]. Figure 2b) shows the corresponding energy spectra of Xe projectiles coincident with recoils. As in Figure 1b), the energies expected for a 40 keV Xe projectile after scattering from hydrogen (38.78 keV), carbon (27.53 keV), and oxygen (24.29 keV) atoms, calculated from classical mechanics, are indicated by dashed lines.

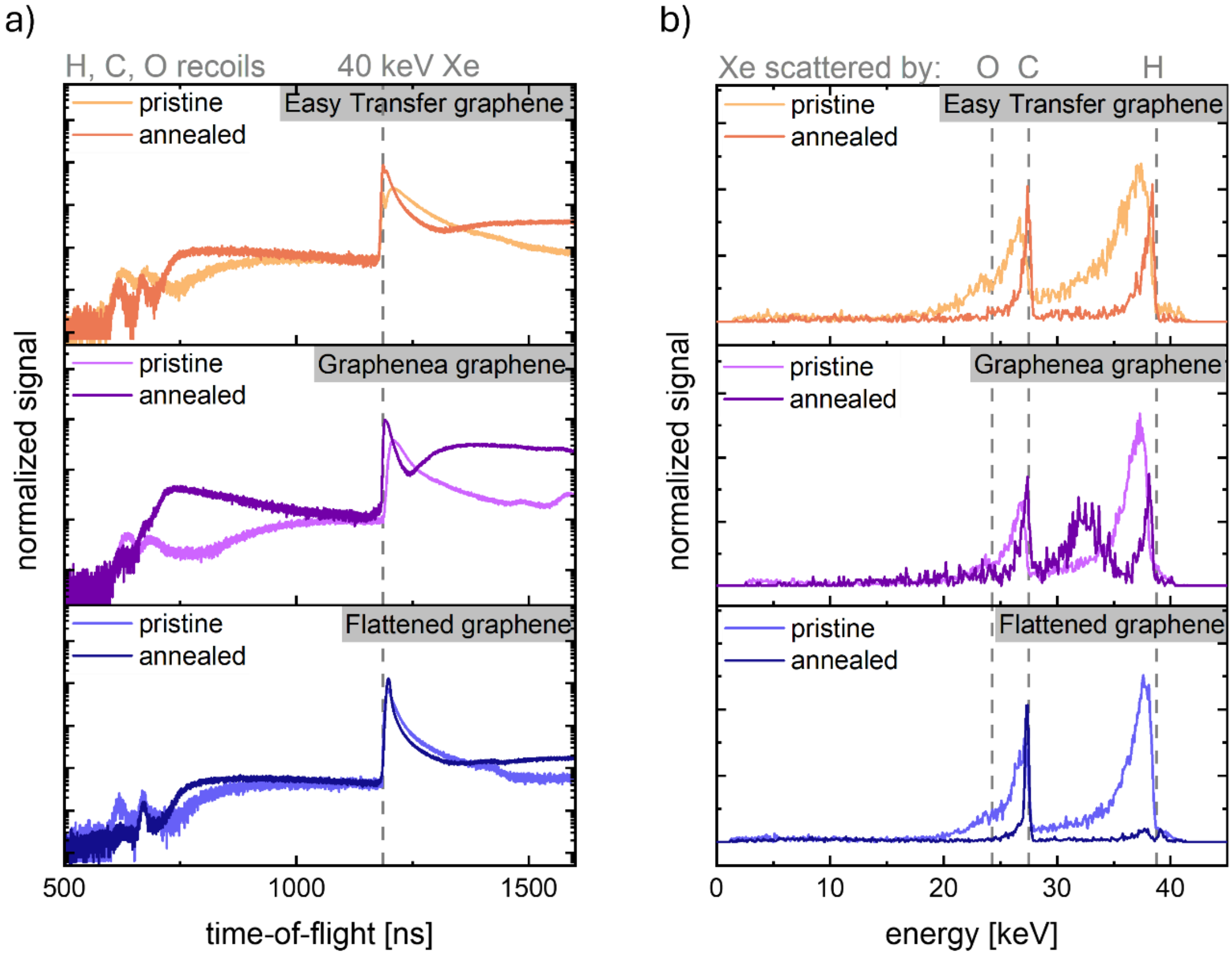

Figure 2: a) Time-of flight spectra recorded for 40 keV Xe ions transmitted through Flattened graphene (bottom, blue, measured 2h 50 min after annealing), Graphenea graphene (middle, purple, measured 2 h 4 min after annealing) and Easy Transfer graphene (top, orange, measured 2 h 16 min after annealing). The measurements of the annealed samples were carried out after cooling to 40 °C. Spectra recorded from the pristine samples are shown in light colors, whereas spectra recorded after annealing at 400 °C for 1 h are shown in darker colors. All spectra are normalized to the number of transmitted Xe. The time-of-flight spectrum of Xe ions transmitted through pristine Flattened graphene is identical to the spectrum shown in Figure 1 of reference [34]. Adapted from reference [34] under the terms of the Creative Commons CC-BY license. b) Corresponding energy spectra of Xe projectiles detected in coincidence with recoils.

In all time-of-flight spectra shown in Figure 2a), the two peaks at short flight times are attributed to hydrogen and carbon recoils originating from the graphene layer, as discussed in Section 5. The signal in the flight time range of approximately 720 ns to 1185 ns is assigned to hydrogen recoils from the Quantifoil support. The flight time of 40 keV Xe ions is indicated by a dashed line. The dominant peak at flight times above approximately 1192 ns corresponds to Xe projectiles transmitted through single-layer graphene.

Annealing significantly modifies the spectra of all three samples. In each time-of-flight spectrum, the hydrogen recoil signal from the graphene layer decreases strongly, while the contribution from the Quantifoil support becomes more prominent, indicating a reduction of the effective thickness of the supporting structure. In the corresponding energy spectra shown in Figure 2b), all samples exhibit a narrowing of the Xe scattering peaks. In all energy spectra, an oxygen edge at 24.4 keV is visible only before annealing, indicating the presence of water and hydroxyl groups. The quantitative carbon and hydrogen coverages obtained from the energy spectra are summarized in Table 1. The error bars represent the relative uncertainty arising from counting statistics.

Besides these general trends, two sample-specific features are observed. For the Graphenea sample, the increased contribution of hydrogen recoils originating from the Quantifoil gives rise to an additional feature at approximately 32 keV in the corresponding energy spectrum.

Table 1: Quantitative carbon and hydrogen coverages for Flattened graphene, Graphenea graphene and Easy Transfer graphene before and approximately 2-3 h after annealing in a $10^{-8}$ mbar vacuum, expressed in monolayer (ML) equivalents. The error bars provide the uncertainty arising from counting statistics.

| **sample** | **C pristine [ML]** | **H pristine [ML]** | **C after cooldown / re-exposure in UHV [ML]** | **H after cooldown / re-exposure in UHV [ML]** | **key observation** |
|---|---|---|---|---|---|
| Flattened graphene | 2.89 ± 0.04 | 2.63 ± 0.09 | 1.04 ± 0.02 | 0.20 ± 0.01 | approaches atomically clean graphene |
| Graphenea graphene | 2.73 ± 0.05 | 3.10 ± 0.04 | 1.41 ± 0.07 | 0.61 ± 0.03 | Quantifoil overlap after annealing |
| Easy Transfer graphene | 4.04 ± 0.06 | 3.63 ± 0.04 | 1.53 ± 0.05 | 0.94 ± 0.03 | highest contamination |

The Flattened graphene sample exhibits the lowest initial total contamination coverage. After annealing, the carbon coverage approaches the 1 monolayer expected for single-layer graphene, and the hydrogen coverage is reduced to only 0.2 monolayers. In contrast, both the Graphenea and Easy Transfer graphene samples exhibit substantially larger carbon and hydrogen coverages after annealing, suggesting both less efficient removal of transfer-related contamination and more pronounced recontamination following annealing. To identify the process, we performed time-resolved surface contamination analysis during sample cooldown. Such measurements are enabled by the minimally destructive character of the recoil-projectile-coincidence approach.

### 2.3 Time-dependent recontamination during cooldown in UHV

In order to investigate carbon and hydrogen surface recontamination during cooldown, the surface coverages of carbon and hydrogen were quantified for Flattened graphene, Easy Transfer graphene, and Graphenea graphene at different times after annealing. Figure 3 summarizes the resulting carbon (grey) and hydrogen (blue) coverages, expressed in monolayer equivalents on the left y-axis as a function of time. The lines between the data points serve as guides to the eye. The error bars correspond to the relative uncertainty arising from counting statistics. The temperature decrease, approximated by a single-exponential decay, is shown as a function of time on the right y-axis function.

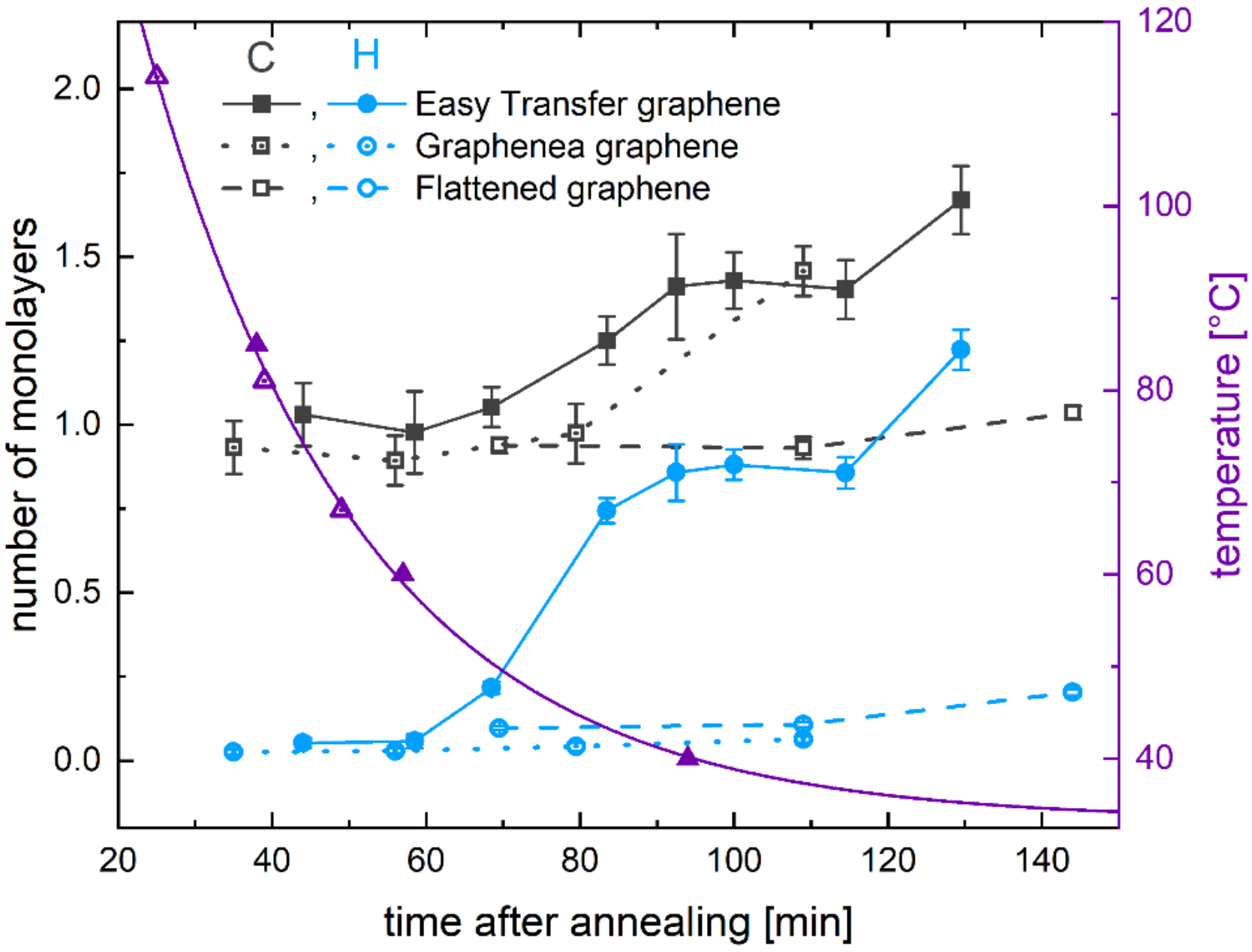

Figure 3: Comparison of carbon (grey) and hydrogen (blue) recontamination during cooling as a function of time of Easy Transfer graphene, Graphenea graphene and Flattened graphene following annealing at 400 °C for 1 h. The lines between the data points serve as a guide to the eye. The error bars correspond to the relative uncertainty arising from counting statistics. The corresponding sample temperature is shown on the right y-axis, approximated with a single-exponential decay ($T(t) = 33.31 + 78.39 \cdot \exp\left[-\frac{t-25.78}{28.06}\right]$).

At the earliest measured time after annealing, all three samples are nearly free of both hydrogen and carbon contamination, with the hydrogen coverage being consistent with zero within the experimental uncertainty, indicating that the subsequently observed hydrogen originates from readsorption during cooldown. The Flattened graphene sample remains essentially unchanged throughout the observation period and shows no significant recontamination even after 140 minutes. In contrast, carbon contamination reappears on the Graphenea graphene sample after approximately 80 minutes, whereas the hydrogen coverage remains close to the detection limit. The recontamination behavior of the Easy Transfer sample differs strongly from that of the other two samples. After approximately 70 minutes, corresponding to a temperature of about 45 °C, both hydrogen and carbon coverages begin to increase rapidly. Between 90 and 115 minutes after annealing, the contamination level remains approximately constant (ca. 1.43 ± 0.08 monolayers C and 0.88 ± 0.05 monolayers H). After 130 minutes, both hydrogen and carbon coverages increase further, suggesting different effective recontamination kinetics, likely influenced by residual-gas adsorption and sample-specific surface chemistry. A typical residual gas spectrum of the vacuum chamber is provided in the Supplementary Information[33].

The differences in contamination levels observed in the ToF-MEIS measurements are also reflected in ex-situ TEM images acquired after annealing. Representative TEM images of Easy Transfer graphene, Graphenea graphene and Flattened graphene are shown in Figures 4a)-c), respectively.

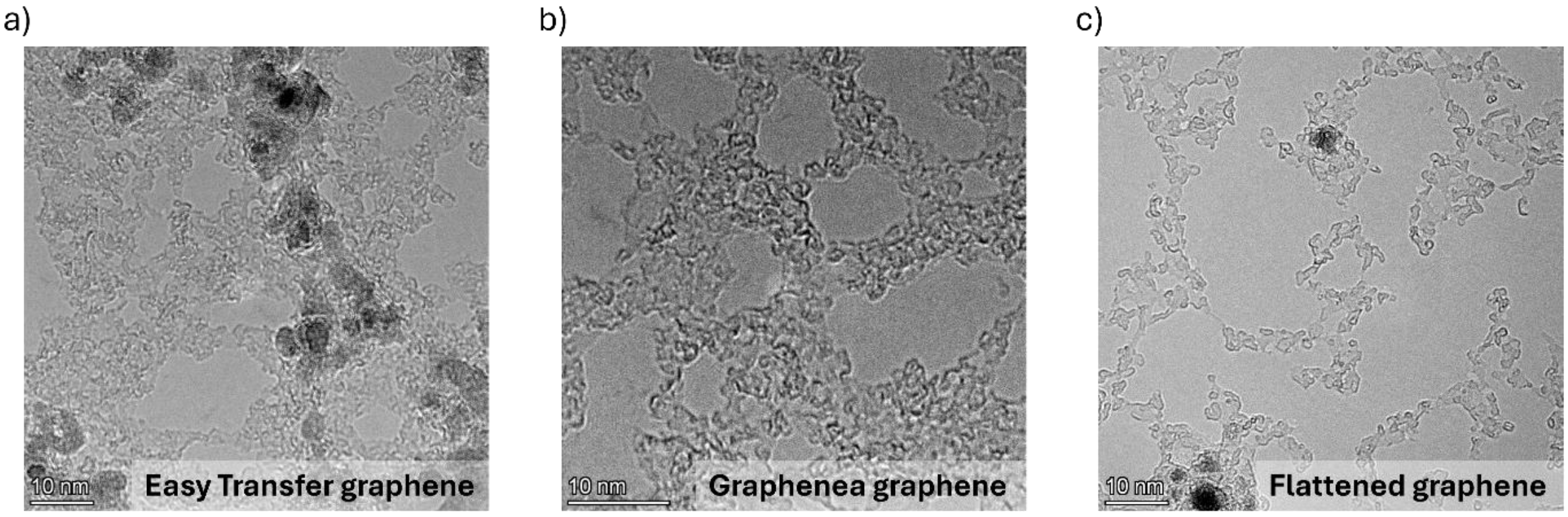

Figure 4: TEM images of a) Easy Transfer graphene, b) Graphenea graphene and c) Flattened graphene recorded ex-situ after annealing. In each panel, the scale bar is 10 nm.

In the TEM images shown in Figure 4, brighter contrast corresponds to cleaner graphene, whereas darker regions indicate visible contamination. All samples exhibit visible contamination clusters; however, the Flattened graphene sample exhibits mostly cleaner areas. In contrast, contaminated regions dominate both Graphenea graphene and Easy Transfer graphene. The cleaner graphene area fractions were estimated by image processing in ImageJ using edge detection, contrast normalization, and automatic thresholding. The resulting cleaner area fractions are approximately 43 % for Easy Transfer graphene, 42 % for Graphenea graphene, and 72 % for Flattened graphene, consistent with the visual impression.

The influence of annealing temperature and duration, as well as the differences in recontamination behaviour among the investigated samples, are discussed in the following section. In addition, we assess the strengths and limitations of the developed coincidence-based method for quantifying surface contamination with isotopic resolution.

## 3. Discussion

The effect of annealing temperature and duration was investigated by ToF-MEIS measurements of PMMA-free transferred Flattened graphene. While hydrocarbon surface contamination was observed on the pristine sample prior to annealing, both hydrogen and carbon surface coverages were found to be significantly reduced by thermal annealing in UHV already at a moderate temperature of 250 °C for 1 h. While the reduction of carbonaceous contamination is consistent with previous STEM results reported by Irschik et al.[27], our measurements additionally provide a direct quantitative determination of the hydrogen coverage, demonstrating that hydrogen contaminants are also efficiently removed during annealing. Notably, extending the annealing duration does not have a significant impact on contaminant removal, as a comparable level of contamination is observed after annealing at 250 °C for 3 h, suggesting that contaminant removal is primarily temperature limited rather than time limited.

After annealing at 400 °C for 1 h, the Flattened graphene sample is found to be nearly free of both carbon and hydrogen contamination, in agreement with the conclusion drawn from the time-of-flight spectrum reported in reference [34]. This result is partially consistent with the findings of Irschik et al., who reported, based on localized STEM measurements, that annealing at 450 °C yielded ca. 70 % relative clean monolayer graphene area, with regions wider than 200 nm[27]. However, while hydrocarbon contamination not directly associated with metal clusters was reported to be removable by annealing at temperatures above 400 °C, metal clusters embedded in carbon-rich contamination were found to limit the achievable cleanliness. In the present study, no such metal clusters are observed.

The observation that the composition of the Flattened graphene sample converges towards one monolayer carbon, as expected for single-layer graphene after annealing at 400 °C for 1 h indicates that the PMMA-free transfer process yields significantly cleaner samples compared to the PMMA-based transfer methods. This conclusion is supported by the direct comparison of the PMMA-free Flattened graphene sample with both PMMA-transferred Graphenea graphene and Easy Transfer graphene, which exhibit significantly higher initial carbon and hydrogen coverages as well as larger contaminant coverages after annealing (section 2.2). These findings are consistent with previous atomic force microscopy (AFM) studies suggesting a higher cleanliness of PMMA-free graphene sample due to the absence of PMMA-residues[22]. In contrast to the localized AFM observations reported previously, the present ToF-MEIS measurements provide element-specific, quantitative determination of both carbon and hydrogen coverages averaged over a macroscopic sample area and enable a direct comparison of contamination levels before and after annealing with isotopic resolution[22].

Although annealing at 400 °C for 1 h results in a removal of surface contamination in all three sample systems immediately after annealing, rapid recontamination is observed for the PMMA-transferred samples, indicating different contamination kinetics. We attribute the recontamination of the graphene surface to adsorption of residual gas species in the UHV environment, suggesting that reduced base pressure extends the duration over which graphene remains clean, in agreement with previous observations of clean PMMA-transferred graphene after annealing in UHV at lower pressure[28]. The delayed recontamination of PMMA-free Flattened graphene is unlikely to be merely a consequence of a lower initial contamination level. Rather, the transfer route appears to define the post-annealing surface state and thus the effective adsorption kinetics. PMMA-assisted transfer may leave residual polymer fragments, chemically active sites, wrinkles, or contamination reservoirs that promote readsorption during cooldown, whereas the PMMA-free Flattened graphene provides fewer nucleation sites for residual-gas adsorption. In addition, PMMA-free Flattened graphene has previously been reported to be less hydrophilic than PMMA-transferred graphene[22], which may further reduce the adsorption of residual gas species.

Our quantitative results on the effect of annealing temperature and duration on graphene cleanliness, as well as on recontamination after annealing, demonstrate the capability of the coincidence-based ToF-MEIS approach. In contrast to local imaging techniques, our method provides element-specific, large-area-averaged quantification of surface contamination with isotopic resolution and monolayer sensitivity while remaining minimally destructive. Since the coincidence approach relies only on ion transmission through sufficiently thin samples, it is directly applicable to a wide range of 2D materials and ultrathin films. This capability enables systematic investigations of adsorption and contamination kinetics, transfer procedures and annealing protocols across different ultrathin systems using the same quantitative method.

Key limitations of this approach include its dependence on scattering cross sections, which are calculated using established interaction potentials (in this study the Universal screening potential implemented in SIMNRA). The corresponding uncertainty contributes systematically to the determined surface coverage but does not affect the relative comparison of samples measured under identical conditions. Furthermore, sufficiently thin targets are required to allow the transmission of ions. Despite these constraints, the precise characterization of the sample surface enables reliable insights into e.g. charge exchange processes in experiments involving projectile transmission through ultrathin films. Our method furthermore provides a platform for extracting effective sticking coefficients for different species under controlled gas-exposure conditions while retaining elemental and isotopic resolution. Extending the coincidence approach to, for example, coincidence-desorption studies could provide additional chemical information.

## 4. Conclusion

We have developed a recoil-projectile coincidence-based, minimally destructive method that enables element-specific identification and quantitative determination of individual surface contaminants in ultrathin materials with isotopic resolution, with a precision primarily limited by counting statistics. In contrast to the predominantly local characterization of graphene cleanliness by TEM or STEM, our approach provides a quantitative, area-averaged determination of surface coverage over macroscopic sample areas, enabling a direct comparison of transfer methods and annealing procedures. Application of this method to self-supporting graphene samples prepared in three different ways reveals that thermal annealing at 400 °C for 1 h under UHV conditions effectively removes carbon and hydrogen surface contamination, yielding large-area-averaged contamination coverages compatible with clean single-layer graphene. We further demonstrate that extending the annealing duration at lower temperatures does not improve contamination removal efficiency. Among the investigated samples, the PMMA-free transferred graphene exhibits the lowest level of initial contamination, approaches the carbon coverage expected for pristine single-layer graphene after annealing at 400 °C and remains clean for at least 140 minutes after annealing at a base pressure of $2 \times 10^{-8}$ mbar. In contrast, contamination due to residual gas adsorption rapidly reappears on the PMMA-transferred samples.

Beyond graphene, the presented coincidence approach establishes a generally applicable quantitative characterization platform for ultrathin materials. Its ability to determine element-specific surface coverages with monolayer sensitivity enables reliable investigations of e.g. surface cleanliness, charge exchange dynamics, sticking coefficients and implanted-species concentrations. Future extensions of the coincidence method, for example combining ion transmission with electron or photon detection, may give access for quantitative probing of quantum mechanical interaction processes at surfaces and in ultrathin materials.

# 5. Methods

## 5.1 Sample systems

Single-layer graphene was transferred onto transmission electron microscopy (TEM) grids with a 10-15 nm thick amorphous carbon support foil (Quantifoil). To investigate the influence of sample preparation on graphene cleanliness, three differently prepared graphene samples, transferred using either a PMMA-based or a PMMA-free process, were studied. The sample systems are denoted as follows:

- **Flattened graphene:** The PMMA-free transfer process described in detail in reference [22] was used. Graphene grown on Cu by CVD was purchased from ACS Material and transferred onto a gold TEM grid. Details of the transfer process are provided in the Supplementary Information[33].
- **Easy Transfer graphene (PMMA)**: A PMMA-assisted transfer process, described in detail in reference [27] was used. Graphene grown on Cu by CVD was purchased from Graphenea and transferred onto a gold TEM grid. Details of the transfer process are provided in the Supplementary Information[33].
- **Graphenea graphene (PMMA)**: Readily transferred graphene on a gold-coated copper TEM grid (PMMA-assisted) purchased from Graphenea.

In each target system, regions with self-supporting single-layer graphene and regions with single-layer graphene supported by Quantifoil are present. In transmission ToF-MEIS, ions transmitted through these regions can be distinguished by their different flight times, as described in Section 5.2. For the present study, the analysis is restricted to the signal originating from ions transmitted through the self-supporting single-layer graphene regions.

## 5.2 ToF-MEIS measurements

Time-of-flight medium energy ion scattering measurements in transmission geometry on self-supporting single-layer graphene were performed at the ToF-MEIS setup at Uppsala University[35,36]. Generated by a 350 kV Danfysik implanter, a beam of 40 keV singly charged Xe ions was transmitted through the target system with the single-layer graphene facing the ion beam, impinging in normal incidence. To reduce pulse lengths and enable precise time-of-flight measurements, the ion beam was electrostatically chopped. Typical pulse lengths of the Xe ion beams used in this study are in the range of 9 ns, corresponding to an energy resolution of approximately 0.3 keV for 40 keV Xe, and the beam spot size on the sample is reduced to a size smaller than 1x1 $mm^2$. To avoid significant damaging of the sample throughout the experiment, the impinging current is kept in the range of one fA.

For a typical measurement ($1 \times 10^7$ incident Xe ions are distributed over an area of $1mm^2$, i.e. at least $10^{13}$ target atoms), the conservative limit of one defect per incident ion corresponds to a maximum defect density of $1 \times 10^9$ $cm^{-2}$, i.e. only 0.26 ppm of the carbon atoms in graphene. The actual defect density is expected to be lower, further supporting the minimally destructive nature of the method. Typical measurement durations in the present study range from 30 to 60 minutes. While shorter acquisition times are generally possible; the achievable statistical uncertainty of the results is determined by counting statistics.

Transmitted projectiles and recoils are detected by a large position-sensitive microchannel plate detector (MCP, DLD120 from RoentDek[37]) located 290 mm from the sample, for the present study centered on the beam axis. In this configuration, the detector covers angles of ±11.5° (solid angle: 0.13 sr). For this study, the evaluated detector region was limited to detection angles of smaller than 7.85°. During the measurements, the base pressure was in the range of $2 \times 10^{-8}$ mbar.

A typical time-of-flight spectrum recorded for 40 keV Xe ions transmitted through pristine Flattened graphene on a gold TEM grid with Quantifoil is shown in Figure 5, including a schematic of the experiment.

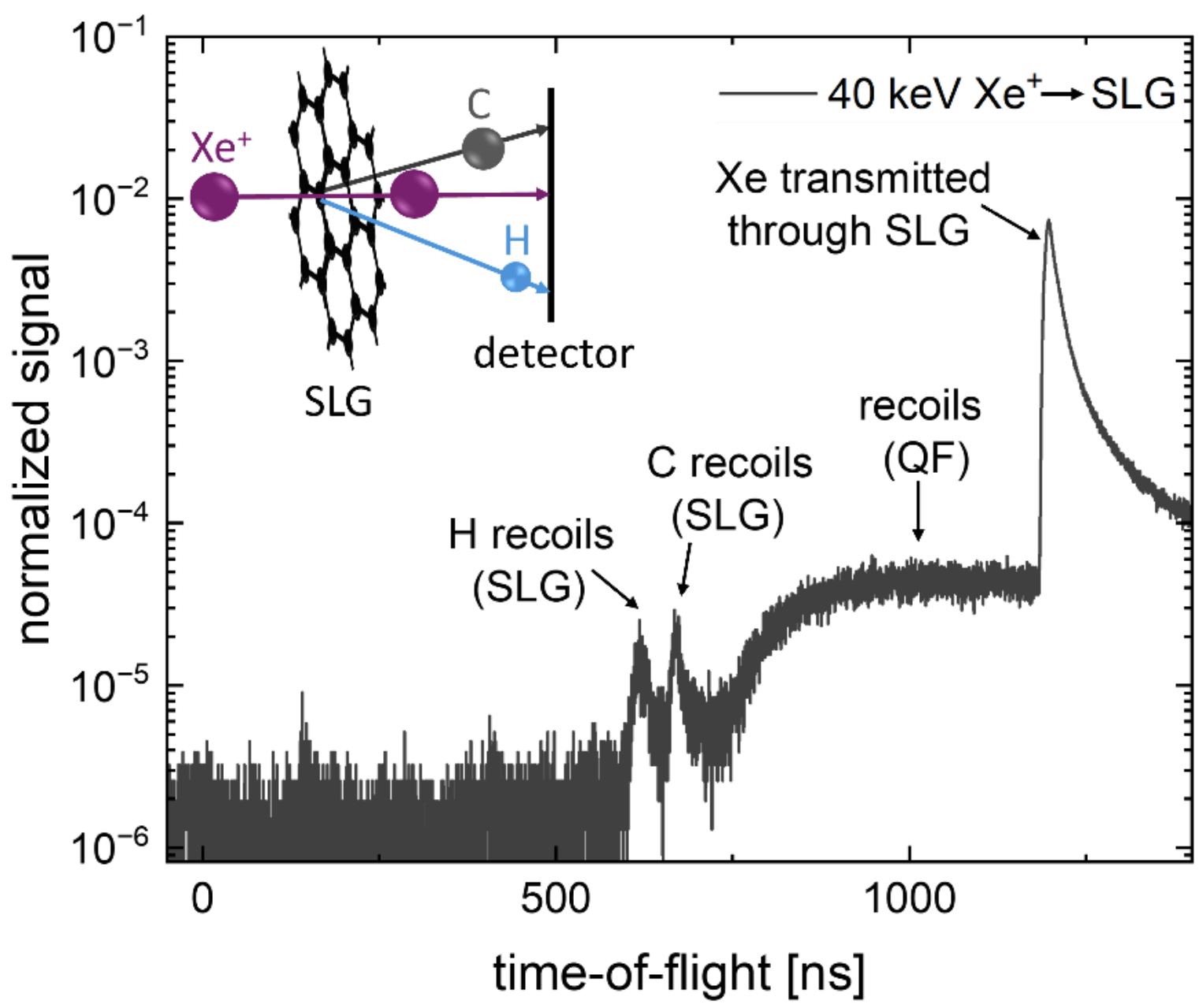

Figure 5: Time-of-flight spectrum of 40 keV Xe ions transmitted through single-layer graphene (SLG), including a schematic of the experiment. Dominant peaks are assigned to H and C recoils from Flattened graphene, to recoils from the Quantifoil (QF) and to Xe projectiles transmitted through graphene.

The two peaks at low flight times, i.e. associated with the fastest particles, are assigned to hydrogen recoils (619 ns) and carbon recoils (671 ns) from the target, respectively. A detailed description of elastic recoil detection analysis in time-of-flight ion scattering is provided in reference[30]. At an energy of 40 keV, the Xe ion beam is transmitted exclusively through regions with self-supporting graphene. Xe ions impinging on a region with Quantifoil are effectively stopped. Xe ions transmitted through pinholes in the graphene, referred to hereafter as the direct beam, are expected at a flight time of 1185.5 ns. For the measurements performed in this study, no significant contribution of the direct beam is observed. The dominant peak at 1197 ns originates from Xe ions transmitted through the single-layer graphene.

### 5.3 Recoil-projectile coincidence condition

To quantify surface contamination and to separate light contamination species with isotopic resolution, a recoil-projectile coincidence condition is defined based on the time-of-flight ranges of recoils and transmitted Xe projectiles. One particle detected by the MCP is required to fall within the time-of-flight interval assigned to H and C recoils from graphene (e.g., 595 ns to 711 ns in Figure 5), while the coincident particle has to be detected within the time-of-flight range corresponding to Xe projectiles transmitted through graphene (e.g., 1185 ns to 1435 ns in Figure 5).

Using this coincidence condition and the MCP effective open area ratio of 54 %, which needs to be considered due to the coincidence condition applied here, the experimentally measured yield $Y_{C,measured}$ can be determined as the ratio of Xe projectiles scattered by carbon atoms ($A_C$) to the number of Xe projectiles transmitted through graphene ($A_{SLG}$):

$$Y_{C,measured} = \frac{A_C}{A_{SLG} \cdot 0.54} .$$

For a carbon monolayer, the expected yield is given by

$$Y_{C,expected} = \frac{2\sigma_{eff}}{A_{UC}},$$

where $\sigma_{eff}$ is the effective scattering cross section and $A_{UC}$ is the area of the graphene unit cell containing two carbon atoms. The differential scattering cross sections $d\sigma/d\Omega$ were calculated using SIMNRA[38] with the Universal screening potential. Since the differential cross sections are approximately constant over the detector acceptance range, the effective scattering cross sections were obtained by multiplying $d\sigma/d\Omega$ by the detector solid angle ($\Omega$ = 0.059 sr). The obtained values are $\sigma_{eff} = 1.87 \times 10^8$ mb for carbon and $\sigma_{eff} = 3.95 \times 10^8$ mb for hydrogen.

The measured number of carbon monolayers, expressed as monolayer equivalents, is obtained from the ratio of the measured and expected yields. The hydrogen coverage is determined analogously using the corresponding recoil counts and scattering cross section. Consequently, the monolayer equivalent $n$ for carbon and hydrogen is given by

$$n_{C/H} = \frac{A_{UC} \cdot A_{C/H}}{2 \cdot \sigma_{eff} \cdot A_{SLG} \cdot 0.54}$$

For atomically clean graphene, a carbon coverage of one monolayer of carbon ($n_C = 1$) and a hydrogen coverage of zero monolayers of hydrogen ($n_H = 0$) are expected.

### 5.4 In-situ annealing experiments

In order to study the effect of annealing on graphene cleanliness, all samples were initially measured in the as-prepared state without undergoing any surface treatment. Subsequent to the pristine measurement, the samples were transferred in-situ to the recently installed preparation chamber[39], where annealing was performed under UHV conditions using a contactless heater. The temperature was monitored using a K-type thermocouple attached to the sample holder.

The influence of both annealing temperature and duration on the cleanliness of graphene was investigated using the Flattened graphene sample. After the initial measurement on the pristine sample, the sample was annealed at 250 °C for 1 h and measured. Subsequently, it was stored in UHV conditions for 10 days to study recontamination before being remeasured and annealed again at 250 °C for 3 h. After being stored in UHV for one day, the sample was measured again and then annealed at 400 °C for 1 h.

For comparison of the different sample transfer processes, both Easy Transfer graphene and Graphenea graphene samples were annealed at 400 °C for 1 h following their pristine measurements.

After each annealing step, the heater current was ramped to zero within 10 minutes and the sample was transferred in-situ to the ToF-MEIS chamber. Measurements were performed both during cooldown and after the sample temperature had decreased below 40 °C.

### 5.5 Ex-situ transmission electron microscopy

Ex-situ transmission electron microscopy was performed after annealing to provide complementary local imaging of the graphene samples. The measurements were carried out using a probe-corrected FEI Titan Themis 200 transmission electron microscope at Uppsala University operated in conventional TEM imaging mode. An acceleration voltage of 80 kV was used to reduce electron-beam-induced damage to graphene. Images were acquired from self-supporting graphene regions suspended over holes in the Quantifoil support film.

For comparison between the different sample systems, representative TEM images were analyzed using ImageJ. The images were processed using contrast normalization, edge detection, and automatic thresholding to estimate the fraction of visibly clean graphene area. The image analysis was used only as a local, semi-quantitative comparison of the spatial distribution of contamination and served as a complement to the large-area, element-specific quantification obtained by ToF-MEIS.

## Acknowledgements

Accelerator operation is supported by the Swedish Research Council VR-RFI (Contract No. 2023-00155). C. F. and M. S. acknowledge financial support from the DFG (CRC 1242 “Non-Equilibrium Dynamics of Condensed Matter in the Time Domain”, project number 278162697, and “2D MATURE”, project number 461605777) and from the Faculty of Physics. E. H. Å. acknowledges financial support from the Research Council of Finland (project number 355011).

## Conflicts of Interests

The authors declare no conflict of interests.

## Data Availability Statement

The data that support the findings of this study are available from the corresponding author upon reasonable request.